\documentstyle[11pt,aaspp4]{article}

\slugcomment{29 Mar 99 draft}

\begin{document}
\title{Red and Blue Shifted Broad Lines in Luminous Quasars
\footnote{Observations reported here were
obtained at the Multiple Mirror Telescope Observatory, a facility operated
jointly by the University of Arizona and the Smithsonian Institution.}}

\author{
D. H. McIntosh\altaffilmark{2},
H.-W. Rix\altaffilmark{3},
M. J. Rieke\altaffilmark{2},
C. B. Foltz\altaffilmark{4}
}

\altaffiltext{2}{Steward Observatory, University of Arizona, Tucson, AZ 85721}
\altaffiltext{3}{Max Planck Institut f\"{u}r Astronomie, Heidelberg, Germany}
\altaffiltext{4}{Multiple Mirror Telescope Observatory, University of Arizona, Tuscon, AZ 85721}

\begin{abstract}
We have observed a sample of 22 luminous quasars, in the range
$2.0 \lesssim z \lesssim 2.5$, at $1.6\micron$ with the near-infrared (NIR) 
spectrograph FSPEC on the Multiple Mirror Telescope.  Our sample contains
13 radio-loud and 9 radio-quiet objects.  We have measured the
systemic redshifts $z_{\rm sys}$ directly from the strong
$[$\ion{O}{3}$]\lambda5007$ line emitted from the narrow-line-region.
From the same spectra, we have
found that the non-resonance broad H$\beta$ lines have a systematic 
mean
{\it redward} shift of $520\pm80$~km~s$^{-1}$ with respect to systemic.
Such a shift was {\it not} found in our identical analysis of the low-redshift
sample of Boroson \& Green.
The amplitude of this redshift is comparable to half the expected
gravitational redshift and transverse Doppler effects, and
is consistent with a correlation between redshift differences and quasar
luminosity.
From data in the literature, we confirm that the high-ionization rest-frame
ultraviolet broad lines are {\it blueshifted}
$\sim 550$--$1050$~km~s$^{-1}$ from systemic, and that these velocity
shifts systematically increase with ionization potential.
Our results allow us to quantify the known bias in
estimating the ionizing flux from the
inter-galactic-medium $J_{\nu}^{\rm IGM}$ via the Proximity Effect.  Using
redshift measurements commonly determined from strong broad line species, like
Ly$\alpha$ or \ion{C}{4}$\lambda1549$, results in an
over-estimation of $J_{\nu}^{\rm IGM}$ by factors of $\sim1.9-2.3$.
Similarly, corresponding lower limits on the density of baryons $\Omega_b$
will be over-estimated by factors of $\sim1.4-1.5$.
However, the low-ionization \ion{Mg}{2}$\lambda2798$ broad line is
within $\sim50$~km~s$^{-1}$ of systemic, and thus would be the 
line of choice for
determining the true redshift of $1.0<z<2.2$ quasars without NIR
spectroscopy, and $z>3.1$ objects using NIR spectroscopy.
\end{abstract}


\section {Introduction}
Defining the center-of-mass (COM $\equiv$ systemic) rest-frame, 
thus $z_{\rm sys}$,
of quasars (QSOs) 
and of their host galaxies is important for a number of reasons.
First, it is crucial for estimating the 
ionizing flux from the
inter-galactic-medium $J_{\nu}^{\rm IGM}$ via the Proximity Effect
(\cite{bajtlik88}).
To estimate the redshift range ($z_{\rm sys} - z_{\rm Ly\alpha}$) over which
Ly$\alpha$ absorbers are underpopulated, we need to know $z_{\rm sys}$
accurately.  An
error of $1000$~km~s$^{-1}$ in $z_{\rm sys}$ leads to a factor of
$\sim2.6$ difference in the inferred $J_{\nu}^{\rm IGM}$ (\cite{espey93}).
In turn, precise estimates of $J_{\nu}^{\rm IGM}$ are needed to
understand the ionization balance of the IGM and the structure of the
Ly$\alpha$ Forest.  Furthermore,
accurate knowledge of the systemic frame is important to understanding
the `central engine' and the broad-line-region (BLR) physics.
Inflow, outflow and
radiative transfer effects may lead to offsets between the emission line
centroids and $z_{\rm sys}$.  The value of $z_{\rm sys}$ also plays
a critical role in understanding the
kinematics of `associated' 
($z_{\rm abs} \approx z_{\rm em}$) absorption line systems (\cite{foltz86})
and the nature of the
`$z_{\rm abs} > z_{\rm em}$' metal line systems (\cite{gaskell82}).
In both cases, determination of $z_{\rm sys}$
is necessary to understand whether the absorbing gas
is in- or out-flowing, or at rest with respect to the QSO.

As Tytler \& Fan (1992) first showed, the QSO redshift as measured 
by different broad emission
lines systematically depends on ionization potential.
In general, it has been well documented (\cite{gaskell82}; \cite{espey89};
\cite{brotherton94}; and references therein) that broad
emission lines from highly ionized species (HILs;
{\it e.g.}~\ion{C}{4}$\lambda1549$, \ion{C}{3}$]\lambda1909$, 
\ion{N}{5}$\lambda1240$), as well as Ly$\alpha$,
are systematically blueshifted
$500$--$1500$~km~s$^{-1}$ with respect to broad emission
lines from lower ionization species (LILs;
{\it e.g.}~\ion{Mg}{2}$\lambda2798$, \ion{O}{1}$\lambda1305$), and the
permitted \ion{H}{1} Balmer series.
In contrast, redshifts
from narrow emission lines 
({\it e.g.} $[$\ion{O}{3}$]\lambda5007$) in low redshift ($z<1$) QSOs 
and Seyfert
galaxies are observed to be within $100$~km~s$^{-1}$ of the broad 
\ion{Mg}{2} and
Balmer lines (\cite{gaskell82}; \cite{tytler92}; \cite{bg92}; \cite{laor95};
and \cite{cb96}).
Furthermore, in local AGN, redshifts from narrow forbidden lines have shown
agreement to $\lesssim100$~km~s$^{-1}$ of the accepted systemic 
frame determined by
stellar absorption features and \ion{H}{1} $21$~cm emission in the host galaxies
(\cite{gaskell82}; \cite{vrtilek85}, \cite{hutchings87}), and are within
$\pm200$~km~s$^{-1}$ of the CO emission (\cite{alloin92}).
Since the narrow emission originates
from a large volume of gas centered on the QSO ({\it i.e.} 
the narrow-line-region with
$r_{\rm NLR}\sim1$~kpc), Penston (1977) first proposed that it is 
likely that the narrow forbidden lines give the COM redshift.

At higher redshifts ($z>1.5$) the strong narrow lines 
can only be detected in the NIR.
Early NIR spectroscopy of a single bright $z=2.09$ QSO showed H$\alpha$, 
H$\beta$ and narrow $[$\ion{O}{3}$]$ emission had similar redshifts to the LILs,
supporting the low redshift studies that claimed Balmer lines provided a 
reliable measure of $z_{\rm sys}$
(\cite{carswell91}).  However, Nishihara {\it et al.} (1997) took NIR spectra
of four $z>1.5$ QSOs and found the broad H$\alpha$ line to be redshifted
on average by $430\pm170$~km~s$^{-1}$ from the systemic $[$\ion{O}{3}$]$
narrow line.  

From our total sample of $32$ bright ($V\leq18.0$ mag), 
$2.0 \lesssim z \lesssim 2.5$ QSOs with NIR $H$-band spectra
(see \cite{mcintosh99}; hereafter M99), we selected
the set of $22$ objects with detected $[$\ion{O}{3}$]\lambda5007$ emission
($[$\ion{O}{3}$]$ EW $\geq 5$\AA, except Q0842+345, Q1011+091, and Q1158-187 for
which the low S/N prohibited reliable line identification).  This
nearly unprecedented detection
of strong emission lines from the NLR of distant,
high luminosity QSOs has provided the direct measurement of $z_{\rm sys}$
via the rest-frame $[$\ion{O}{3}$]\lambda5007$ line center.  Consequently,
this unique data set allows
direct comparisons between the COM frame
and a variety of resonance and recombination broad line centers in this sample.
In addition, each spectrum contains the broad
H$\beta$ line, thus, we are able to directly compare the NLR
redshift with a redshift obtained from BLR emission in
the same spectrum.

\section {Emission Line Redshift Measurements}

To measure $z_{\rm sys}$ for
each QSO, we
fit roughly the upper half of the $[$\ion{O}{3}$]\lambda5007$ 
line with a Gaussian, except for three objects (Q1228+077, Q1331+170
and Q1416+091) for which the peak or centroid of the
$[$\ion{O}{3}$]$ line was used due to the poor S/N of these spectra.  This
fitting procedure provided line center measurements with much smaller
uncertainties compared with the broad H$\beta$ measurements (see below).
We estimated the uncertainties to be $\sigma \sim \pm50$~km~s$^{-1}$.
The measured systemic redshifts are presented in column (2) of Table~1.

To determine the QSO broad H$\beta$ redshift $z_{\rm em}$, we shifted
each spectrum to its systemic frame
and fit it with a set of templates 
representing $[$\ion{O}{3}$]$, H$\beta$, \ion{Fe}{2} and continuum emission 
features using a $\chi^2$ optimization routine (as described in M99).  
All components except
H$\beta$ were subtracted from the data spectrum resulting in an H$\beta$ line 
spectrum.  The central wavelength of the broad component was 
obtained by fitting a single Gaussian to all portions of
the line profile with
relative flux $\leq \case{3}{4}$ of the maximum (see M99).  The 
broad H$\beta$ line centers measured in the systemic frame, with
their $\pm1\sigma$ uncertainties, are given
in column (3) of Table~1.  In addition, the emission redshifts derived from 
broad H$\beta$ are listed in column (4).

\clearpage
\begin{scriptsize}
\begin{deluxetable}{lcccr}
\tablewidth{0pt}
\tablenum{1}
\tablecaption{Observed~Redshift~Parameters}
\tablehead{\multicolumn{1}{l}{QSO} & \colhead{$z_{\rm sys}$} & \multicolumn{3}{c}{Broad~H$\beta$} \\
\cline{3-5} \\
\colhead{} & \colhead{} & \colhead{$\lambda_{\rm ctr}$} & \colhead{$z_{\rm em}$} & \colhead{$\Delta v$}\\
\colhead{} & \colhead{} & \colhead{(\AA)} & \colhead{} & \colhead{(km ${\rm s}^{-1}$)}
}
\startdata
BAL0043+008 & 2.146 & $4866\pm2$ & 2.149 & $310\pm120$ \nl
Q0049+014 & 2.307 & $4862\pm4$ & 2.308 & $60\pm310$ \nl
Q0109+022 & 2.351 & $4868\pm6$ & 2.356 & $430\pm370$ \nl
Q0123+257 & 2.370 & $4868\pm3$ & 2.375 & $430\pm190$ \nl
Q0153+744 & 2.341 & $4871\pm2$ & 2.348 & $620\pm190$ \nl
Q0226-038 & 2.073 & $4858\pm5$ & 2.071 & $-190\pm310$ \nl
Q0421+019 & 2.056 & $4873\pm3$ & 2.064 & $740\pm190$ \nl
Q0424-131 & 2.168 & $4866\pm3$ & 2.171 & $310\pm190$ \nl
Q0552+398 & 2.363 & $4867\pm3$ & 2.367 & $370\pm190$ \nl
HE1104-181 & 2.318 & $4870\pm3$ & 2.324 & $560\pm190$ \nl
Q1148-001 & 1.980\tablenotemark{a} & \nodata & \nodata & \nodata \nl
Q1222+228 & 2.058 & $4858\pm3$ & 2.056 & $-190\pm190$ \nl
Q1228+077 & 2.389 & $4860\pm3$ & 2.388 & $-60\pm190$ \nl
PG1247+267 & 2.042 & $4873\pm1$ & 2.050 & $740\pm\;\:60$ \nl
Q1331+170 & 2.097 & $4861\pm4$ & 2.097 & $0\pm250$ \nl
Q1416+091 & 2.017 & $4851\pm8$ & 2.011 & $-620\pm490$ \nl
Q1435+638 & 2.066 & $4875\pm4$ & 2.075 & $860\pm250$ \nl
Q1448-232 & 2.220 & $4876\pm2$ & 2.230 & $930\pm120$ \nl
Q1704+710 & 2.010 & $4879\pm5$ & 2.021 & $1110\pm310$ \nl
BAL2212-179 & 2.228 & $4864\pm5$ & 2.230 & $190\pm310$ \nl
Q2251+244 & 2.359 & $4852\pm4$ & 2.353 & $-560\pm250$ \nl
Q2310+385 & 2.181 & $4869\pm4$ & 2.186 & $490\pm250$ \nl
\enddata
\tablenotetext{a}{Redshift of this QSO is such that H$\beta$ line center is
blueward of $H$-band window.}
\end{deluxetable}
\end{scriptsize}
\clearpage

\section {Velocity Shifts}
Comparing the $z_{\rm sys}$ and the broad H$\beta$ $z_{\rm em}$ measurements
for each QSO (see Table~1) reveals a
velocity shift $\Delta v$ relative to the systemic frame.
A $\Delta v>0$ corresponds to a redward shift of
H$\beta$ with respect to systemic, while a negative $\Delta v$ 
represents a blueshift.  The velocity shifts for broad H$\beta$ are tabulated
in column (5) of Table~1, where the uncertainties are determined by the
broad H$\beta$ component.

This is the first sizeable sample of high luminosity
QSOs to permit a statistical
comparison of the velocity shifts of various UV rest-frame broad line
redshifts relative to {\it systemic}.  Therefore, we compiled the published
redshifts for each object in Table 2, and calculated $\Delta v$ for each line
species:
\begin{equation}
\Delta v = \left( \frac{z_{\rm line}-z_{\rm sys}}{1+z_{\rm sys}} \right) c  ,
\end{equation}
where $z_{\rm line}$ is the published redshift.
The distribution of velocity shifts for each UV line species
is plotted separately in Figure~1.  All blueshifts larger than
$2250$~km~s$^{-1}$ are due to a single radio-loud object (Q2251+244) with
strong $[$\ion{O}{3}$]$ emission (see M99, Fig. 1).
The $z_{\rm sys}$ measurements are accurate to 
$\sim50$~km~s$^{-1}$ on average, while the majority of the published
redshifts are known to $\pm0.001$ which corresponds to roughly 
$\pm100$~km~s$^{-1}$.  

\clearpage
\begin{tiny}
\begin{deluxetable}{lcccccccccccc}
\tablewidth{0pt}
\tablenum{2}
\tablecaption{Published~Emission~Line~Redshifts}
\tablehead{\multicolumn{1}{l}{QSO} & \colhead{Ly$\alpha$} & \colhead{Refs.\tablenotemark{a}} & \colhead{\ion{N}{5}} & \colhead{Refs.\tablenotemark{a}} & \colhead{\ion{Si}{4}/\ion{O}{4}$]$\tablenotemark{b}} & \colhead{Refs.\tablenotemark{a}} & \colhead{\ion{C}{4}} & \colhead{Refs.\tablenotemark{a}} & \colhead{\ion{C}{3}$]$} & \colhead{Refs.\tablenotemark{a}} & \colhead{\ion{Mg}{2}} & \colhead{Refs.\tablenotemark{a}
}\\
\colhead{} & \colhead{$\lambda1216$} & \colhead{} & \colhead{$\lambda1240$} & \colhead{} & \colhead{$\lambda1400$} & \colhead{} & \colhead{$\lambda1549$} & \colhead{} & \colhead{$\lambda1909$} & \colhead{} & \colhead{$\lambda2798$
}}
\startdata
BAL0043+008 & \nodata & \nodata & \nodata & \nodata & \nodata & \nodata & 2.141 & 1,2 & 2.143 & 2,3 & 2.143 & 3\tablenotemark{c} \nl
Q0049+014 & 2.31 & 4 & \nodata & \nodata & \nodata & \nodata & 2.285 & 5 & 2.30 & 6\tablenotemark{d} & \nodata & \nodata \nl
Q0109+022 & 2.35 & 4 & \nodata & \nodata & \nodata & \nodata & 2.348 & 5 & 2.34 & 6\tablenotemark{d} & \nodata & \nodata \nl
Q0123+257 & 2.360 & 7,8\tablenotemark{d} & \nodata & \nodata & \nodata & \nodata & 2.356 & 7,8\tablenotemark{d} & 2.36 & 6\tablenotemark{d} & \nodata & \nodata \nl
Q0153+744 & 2.342 & 9 & 2.340 & 9 & 2.340 & 9 & 2.339 & 9 & 2.339 & 9 & 2.349 & 9 \nl
Q0226-038 & 2.069 & 10-12 & 2.058 & 12 & 2.057 & 10,12,13 & 2.064 & 10-14 & 2.064 & 14 & 2.073 & 14 \nl
Q0421+019 & 2.048 & 12,15 & \nodata & \nodata & \nodata & \nodata & 2.051 & 15 & 2.053 & 12,14 & 2.056 & 14 \nl
Q0424-131 & 2.160 & 13,16 & \nodata & \nodata & 2.168 & 13 & 2.164 & 13,14,16 & 2.164 & 14,16 & 2.166 & 14,16 \nl
Q0552+398 & 2.362 & 4,17 & \nodata & \nodata & \nodata & \nodata & \nodata & \nodata & \nodata & \nodata & \nodata & \nodata \nl
HE1104-181 & 2.315 & 18 & \nodata & \nodata & \nodata & \nodata & 2.312 & 18-20 & \nodata & \nodata & \nodata & \nodata \nl
Q1148-001 & 1.979 & 11,12 & \nodata & \nodata & 1.975 & 10,11,12 & 1.979 & 10-14,21 & 1.979 & 10,14,21 & 1.978 & 14,21 \nl
Q1222+228 & 2.046 & 11,12,16 & 2.049 & 12 & 2.046 & 12,13 & 2.042 & 11-13,16 & 2.053 & 16 & 2.056 & 16 \nl
Q1228+077 & 2.388 & 4,22,23 & \nodata & \nodata & \nodata & \nodata & 2.389 & 24 & 2.392 & 12 & \nodata & \nodata \nl
PG1247+267 & 2.040 & 11-13 & 2.042 & 12 & 2.040 & 12,13 & 2.039 & 11-13 & 2.043 & 25\tablenotemark{d} & \nodata & \nodata \nl
Q1331+170 & 2.084 & 11,12,16,26 & 2.079 & 12,26 & 2.086 & 12,13 & 2.079 & 11-13,16,26 & 2.086 & 16,26 & 2.095 & 16,26 \nl
Q1416+091 & \nodata & \nodata & \nodata & \nodata & \nodata & \nodata & \nodata & \nodata & 2.012 & 14 & 2.018 & 14 \nl
Q1435+638 & \nodata & \nodata & \nodata & \nodata & 2.072 & 13 & 2.066 & 13,14 & 2.062 & 14,27 & 2.061 & 14 \nl
Q1448-232 & 2.216 & 16,28 & \nodata & \nodata & \nodata & \nodata & 2.216 & 16,28,29 & 2.214 & 16,29 & 2.223 & 16 \nl
Q1704+710 & \nodata & \nodata & \nodata & \nodata & \nodata & \nodata & \nodata & \nodata & 2.011 & 14 & 2.019 & 14 \nl
BAL2212-179 & 2.220 & 30\tablenotemark{e} & 2.205 & 30\tablenotemark{e} & 2.218 & 30\tablenotemark{e} & 2.209 & 30\tablenotemark{e} & \nodata & \nodata & \nodata & \nodata \nl
Q2251+244 & 2.330 & 7,10,31 & \nodata & \nodata & 2.328 & 10,12 & 2.323 & 7,10,12,31 & 2.318 & 12,31 & \nodata & \nodata \nl
Q2310+385 & \nodata & \nodata & \nodata & \nodata & \nodata & \nodata & 2.175 & 32 & \nodata & \nodata & \nodata & \nodata \nl
\enddata
\tablenotetext{a}{Multiple references indicate that the redshift is an average.}
\tablenotetext{b}{$\lambda1400$ blend of the \ion{Si}{4} doublet and the \ion{O}{4} multiplet.}
\tablenotetext{c}{Redshift extrapolated from an emission line profile comparison.}
\tablenotetext{d}{Redshift measured directly from published spectrum.}
\tablenotetext{e}{Redshift measured directly from digital spectrum.}
\tablerefs{(1) Osmer {\it et al.} 1994; (2) Turnshek {\it et al.} 1980; (3) Hartig \& Baldwin 1986; (4) Wolfe {\it et al.} 1986; (5) Schneider {\it et al.} 1994; (6) Pei {\it et al.} 1991; (7) Schmidt 1975; (8) Barlow \& Sargent 1997; (9) Lawrence {\it et al.} 1996; (10) Gaskell 1982; (11) Young {\it et al.} 1982; (12) Tytler \& Fan 1992; (13) Sargent {\it et al.} 1988; (14) Steidel \& Sargent 1991; (15) Schmidt 1977; (16) Espey {\it et al.} 1989; (17) Wills \& Wills 1976; (18) Smette {\it et al.} 1995; (19) Wisotzki {\it et al.} 1993; (20) Reimers {\it et al.} 1996; (21) Aldcroft {\it et al.} 1994; (22) Vaucher \& Weedman 1980; (23) Robertson \& Shaver 1983; (24) Sramek \& Weedman 1978; (25) Green {\it et al.} 1980; (26) Carswell {\it et al.} 1991; (27) Laor {\it et al.} 1995; (28) Jian-sheng {\it et al.} 1984; (29) Ulrich 1989; (30) Morris {\it et al.} 1991; (31) Barthel {\it et al.} 1990; (32) B. J. Wills 1997, private communication}
\end{deluxetable}
\end{tiny}
\clearpage

\section {Results and Discussion}
From the above analysis, we find:

\begin{enumerate}
\item that the H$\beta$ broad component emission has a weighted
mean redshift of $520\pm80$~km~s$^{-1}$ relative to
$z_{\rm sys}$, with a sample variance of $360$~km~s$^{-1}$.

\item that on average the \ion{Mg}{2}$\lambda2798$ based redshift
is within $\sim 50$~km~s$^{-1}$ of systemic,
confirming that LILs should
provide reliable $z_{\rm sys}$ estimates (\cite{carswell91}).
The rest wavelength and strength of \ion{Mg}{2} emission make it 
the line of choice for
determining $z_{\rm sys}$ of distant ($1.0<z<2.2$) QSOs without
NIR spectroscopy, or even more remote ($z\gtrsim3.1$) objects
using NIR spectroscopy.

\item confirmation of the following effects: 
(i) The average blueshift ($\sim 600-1100$~km~s$^{-1}$) of the HILs
relative
to the LIL \ion{Mg}{2}.
(ii) The correlation between average blueward
velocity shift
and ionization potential for each resonance broad line species.

\item no correlation between QSO radio type ({\it i.e.} radio-loud versus
radio-quiet) and any of the
line species velocity shifts.
\end{enumerate}

\subsection {Redshifted Balmer Lines}
The mean redshift of broad H$\beta$
is similar to that found for H$\alpha$ in a small sample (4) of luminous
$z\sim2$ QSOs (\cite{nishihara97}).
Yet at lower redshifts ($z<1$), 
Balmer emission gives
the same redshift to within $100$~km~s$^{-1}$ of narrow forbidden emission.  
Given that $z\sim2$ QSOs are more luminous than their present-day counterparts,
perhaps this
observed trend of increased Balmer redshift with increased $z_{\rm sys}$
represents a luminosity dependence.
In fact, we find a correlation between H$\beta$ redshift
and QSO rest-frame $V$-band luminosity (from
M99), such that the $\sim100$
times less luminous sample of Boroson \& Green (1992) exhibits 
a mean broad H$\beta$ redshift of $<100$~km~s$^{-1}$
(see Figure~2).  We conclude that
the use of Balmer emission lines to measure $z_{\rm sys}$ may not be
prudent for more distant, and thus more luminous QSOs.

The magnitude of the redward Balmer shift may be partially due to
relativistic effects (\cite{netzer77}).
Recent studies ({\it e.g.} \cite{cb96}) have suggested a connection
between the luminosity dependence of QSO broad line redward asymmetries
and gravitational redshifts due to virialized motions 
near super-massive ($\sim10^9$--$10^{10}~M_{\sun}$) black
holes.  Assuming BLR emission from a gravitationally bound
(virialized), spherical ensemble of clouds at $r\sim0.1$~pc 
allows the conversion
from the observed mean broad line width to the 
mean line-of-sight (LOS) velocity and the true average cloud velocity:
$\langle v_{\rm FWHM} \rangle = 2 \langle v_{\rm LOS} \rangle = 
\frac{2}{\sqrt{3}} \langle v_{\rm cloud} \rangle$.
We measured the weighted mean line width 
$\langle v_{\rm FWHM} \rangle = 8750\pm570$~km~s$^{-1}$, with a sample
variance of $2550$~km~s$^{-1}$, from the H$\beta$
broad component fitting (M99).
The gravitational redshifting of emission from clouds
deep within the potential well of the super-massive black hole is
$\Delta v_{\rm GR} \simeq \frac{\langle v^{2}_{\rm cloud} \rangle}{c} \simeq 190\pm20 \; {\rm km} \; {\rm s}^{-1}$.
In addition, emission from clouds moving perpendicular to our LOS experiences a 
transverse Doppler shift of
$\Delta v_{\rm TDS} \simeq \frac{v^{2}_{\bot}}{2c} \simeq \frac{\langle v^{2}_{\rm cloud} \rangle}{3c} \simeq 60\pm10 \; {\rm km} \; {\rm s}^{-1}$.

The combination of these two relativistic effects can account for roughly half
of the $520\pm80$~km~s$^{-1}$ Balmer redshift.  Although intriguing,
it is unclear why this emission would not also
exhibit Doppler beaming effects ({\it e.g.} \cite{laor91}) and why
the HILs, which reverberation mapping studies have
placed at the same distance, have blueshifts.

\subsection {Cosmological Implications of the $z_{\rm sys}$ Measurements}
HILs, like Ly$\alpha$ or \ion{C}{4}, are not directly
measuring $z_{\rm sys}$.  Specifically, for this sample we find Ly$\alpha$ 
and \ion{C}{4} to be blueshifted on average $550$ and $860$~km~s$^{-1}$,
respectively, relative to systemic.  Therefore, using redshifts acquired from 
these lines will result in the
over-estimation of $J_{\nu}^{\rm IGM}$ via Proximity Effect calculations by
respective factors of roughly 1.9 and 2.3 (\cite{espey93}).
In addition, the lower limit to the
density of baryonic matter in the Universe $\Omega_b$, is proportional to the
square root of $J_{\nu}^{\rm IGM}$ (\cite{rauch97}).  Therefore,
over-estimating the strength of the background flux translates into an
over-estimation of the lower limit of $\Omega_b$ by factors of about $1.4$ 
and $1.5$.

\acknowledgments{}
We are grateful to Ray Weymann who generously reviewed this Letter and provided
insightful comments and suggestions.
We acknowledge helpful discussions with 
Robert Antonucci, Jill Bechtold, John Cocke, Mike Corbin, Rolf Kudritzki,
Phil Pinto, Jennifer Scott, Matthias Steinmetz and Belinda Wilkes.  And we
thank the anonymous referee for fair and thoughtful suggestions that 
improved our Letter.  This research
was supported, in part, by National Science Foundation Grants AST 9529190 and
9803072.

\clearpage


\newpage


\begin{figure}[p]
\centering
\plotone{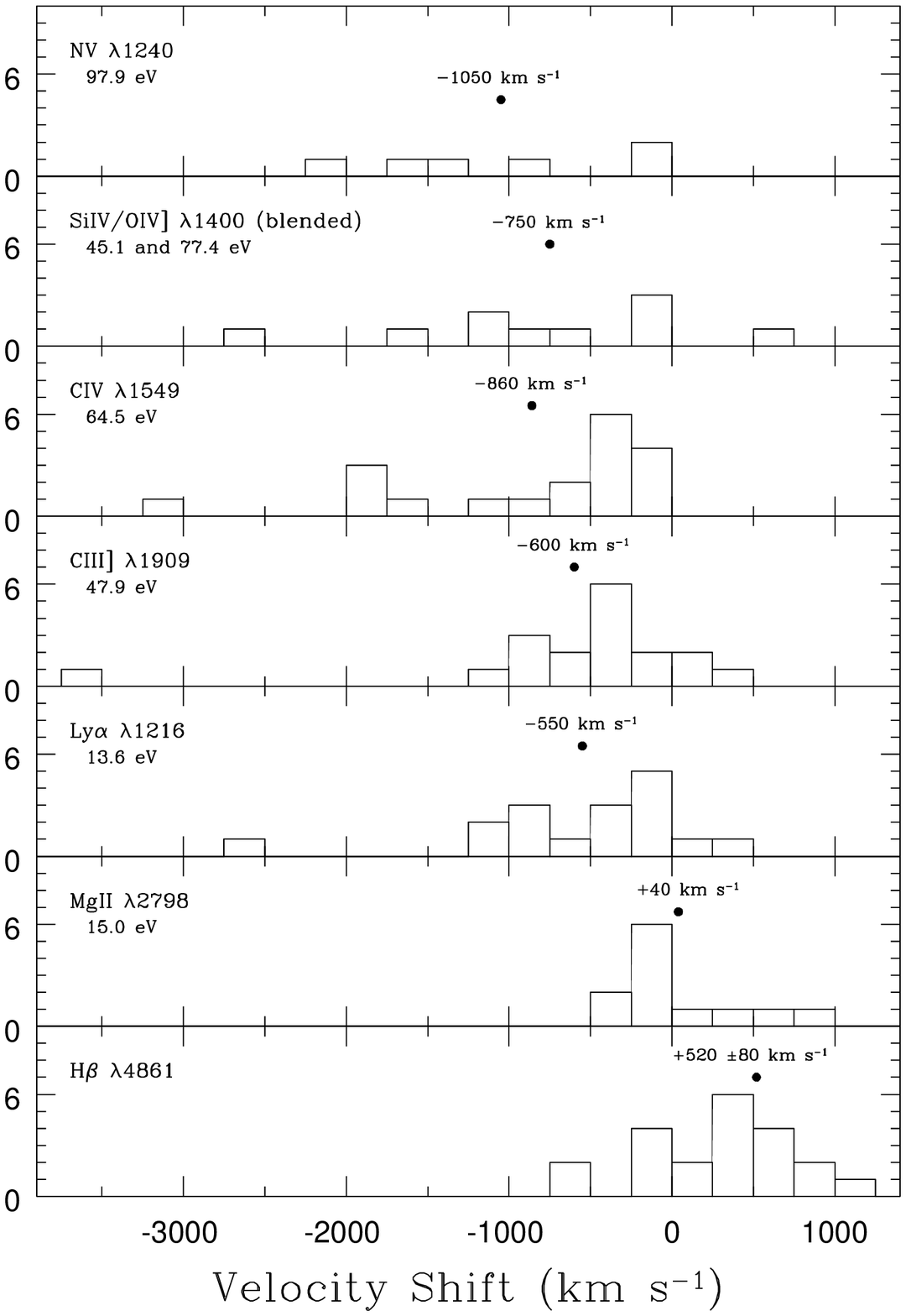}
\figcaption{Distributions of the broad emission
line velocity shifts relative to the systemic frame (given by
the $[$\ion{O}{3}$]\lambda5007$ emission line center).  A negative velocity
shift corresponds to a blueshift relative to systemic, while a positive
shift represents a redshift.  The velocity bins are
$\sim 1\langle \sigma \rangle$ for the broad H$\beta$ shifts and 
$\sim 2.5\langle \sigma \rangle$ for the UV rest-frame line shifts.  The
unweighted average of each blueshifted distribution is plotted as a solid dot,
while the weighted mean with $\pm 1\sigma$ uncertainty is shown for the
redshifted broad H$\beta$ distribution.  The ionization potential 
$\chi_{\rm ion}$ 
of each UV line is given, as well as its rest wavelength (in \AA).}
\end{figure}

\begin{figure}[p]
\centering
\plotone{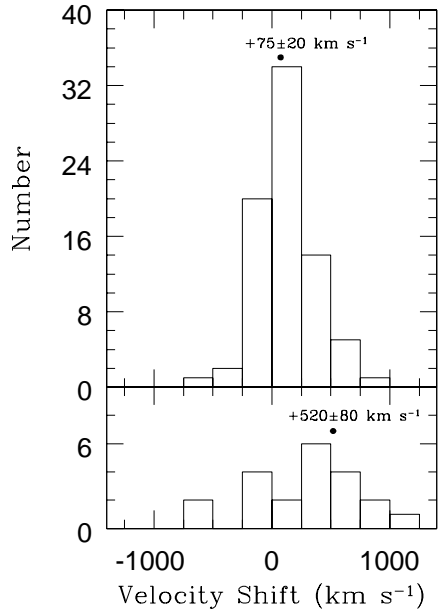}
\figcaption{The comparison
between the distributions of broad H$\beta$ velocity shifts for
the nearby ($z<0.5$, upper panel), lower luminosity
sample of Boroson \& Green (1992); and our $\sim 100$ times 
more luminous, high redshift
($z>2.0$, lower panel) sample.  The broad component H$\beta$ line centers,
used to calculate the velocity shifts for each sample, were measured using
the identical procedure (described in $\S 2.2$) and 
$[$\ion{O}{3}$]$ EW $\geq 5$\AA $\,$ selection criteria.
The mean rest-frame $V$-band luminosities of these two
samples differ by roughly a factor of $100$ (see M99).  The data bin size
is $\sim 1\langle \sigma \rangle$ of the broad H$\beta$ velocity shifts found 
in the more luminous sample.  The mean of each distribution is plotted 
as a solid dot.}
\end{figure}
 
\end{document}